\documentstyle[11pt,paspconf,epsf]{article}
\input epsf

\begin{document}
\title{On the QSO content of the First Byurakan Survey and the completeness of 
the Palomar Green Survey} 

\author{P. V\'eron\altaffilmark{1}, A.\,M. Mickaelian\altaffilmark{2},  
A.\,C. Gon\c{c}alves\altaffilmark{1} and M.-P. V\'eron-Cetty\altaffilmark{1} }

\affil{ $^{1}$\,Centre National de la Recherche Scientifique, 
Observatoire de Haute Provence, 04870 St. Michel l'Observatoire, 
France}
\affil{ $^{2}$\,Byurakan Astrophysical Observatory, Byurakan 378433, 
Republic of Armenia}

\begin{abstract}
The second part of the First Byurakan Survey is aimed at detecting all bright
($B <$ 16.5) UV-excess starlike objects in a large area of the sky. By comparison 
with major X-ray and radio surveys we tentatively identified as QSOs 11 FBS 
objects. We made spectroscopic observations of nine of them. We found six new 
QSOs bringing the total number of known QSOs in this survey to 42.

By comparison with the Palomar Green (PG) QSO survey, we found that the 
completeness of this last survey is of the order of 70\% rather than 
30--50\% as suggested by several authors.
\end{abstract}

\keywords{Quasars -- Surveys}

\vspace*{5mm}
\section{Introduction}

The surface density of bright QSOs ($B <$ 17.0) is still very poorly known. The 
PG survey (Schmidt \& Green 1983) covering an area of 10\,714 
deg$^{2}$ lead to the discovery of 69 QSOs brighter than $M_{B}$ 
= $-$24 ($H_{\rm o}$ = 50 km\,s$^{-1}$\,Mpc$^{-1}$) and $B$ = 16.16, 
corresponding to 0.0064 deg$^{-2}$. However, several authors have concluded 
that it could be incomplete by a factor 2 to 3. 
  
The First Byurakan Survey (FBS), also known as the Markarian survey, was 
carried out in 1965--80 by Markarian et al. (1989). It covers 17\,000 
deg$^{2}$ 
and is complete to about $B$ = 16.5$^{\rm m}$. It has been 
used by Markarian and his 
collaborators to search for UV excess galaxies; it can also be used for finding
UV-excess starlike objects. Such a program --- the second part 
of the FBS --- has been undertaken in 1987 (Abrahamian et al. 1990; Mickaelian 
1994). Its main purpose is to take advantage of the large sky area covered to 
get a reliable estimate of the surface density of bright QSOs. 

At the present time, 4\,109 deg$^{2}$ have been searched and a 
catalogue of 1\,103 blue stellar objects (715 at 
$\vert b \vert >$ 30$^{\rm o}$) has been built. 434 
spectroscopic identifications (398 stars and 36 QSOs) are already 
known. Photoelectric UBV photometry for 113 FBS objects has been 
published. For all except one, $U - B$ $<$ $-$0.50.

\section{Comparison with X-ray and radiosurveys}

\subsection{The ROSAT All-sky Survey Bright Source 
Catalogue (RASS-BSC)}

We have cross-correlated the list of 1\,103 UV-excess objects with the 
RASS-BSC (Voges et al. 1996). We 
have found seven coincidences with FBS objects of unknown nature.

\subsection{The WGA ROSAT catalogue of point sources (WGACAT)}

The WGACAT has been generated using the {\it ROSAT} PSPC pointed data 
publicly available as of September 1994. It contains more than 45\,600 
individual sources (White et al. 1994). We have found two coincidences 
with FBS objects of unknown nature.

\subsection{The NRAO VLA sky survey (NVSS)}

The NVSS covers the sky north $\delta$ = $-$40$^{\rm o}$ at 1.4 GHz 
(Condon et al. 1998). It contains almost 2$\times$10$^{6}$ discrete sources 
stronger than S = 2.5 mJy. We have found two coincidences with 
FBS objects of unknown nature.

\section{Observations}

Spectroscopic observations of nine of the eleven objects coincident either 
with a X-ray or a radio source were carried out with the Observatoire de 
Haute Provence 1.93 m telescope. 
Six (listed in Table 1; the spectra are shown in Fig. 1) turned out to 
be QSOs, while three are stars. 

\section{Discussion}
\subsection{Completeness of the FBS}

The Catalogue of mean UBV data on stars (Mermilliod \& Mermilliod 1994) 
contains 102 stars in the FBS area, with 11.0 $< V <$ 16.5 (bright stars are 
saturated on the FBS plates and are therefore missed) and $U - B$ $<$ 
$-$0.50; 53 are included in the FBS catalogue suggesting that, in this 
magnitude interval, the completeness of the FBS is 52\% (53/102). 
The completeness of the survey 
increases from $\sim$ 20\% at $U - B$ $\sim$ $-$0.6 to $\sim$ 80\% 
for $U - B$ $<$ $-$1.0. 
The QSO $U - B$ colour changes with $z$; but most of the changes, at least for 
$z$ $<$ 2.2, are due to the presence of an emission line in one of 
the two filters. The $U - B$ colour of the continuum is in the range 
$-$0.9 $<$ $U - B$ $<$ $-$0.7. Slitless spectroscopic surveys are 
sensitive to the colour of the continuum, unaffected by the emission 
lines. There are 58 known stars with 14.0 $< V <$ 16.5 and 
$U - B$ $<$ $-$0.70 in the FBS area; 39 (67\%) have been found by the 
FBS; we shall adopt this value for the completeness of the FBS.

\begin{figure}
\plotfiddle{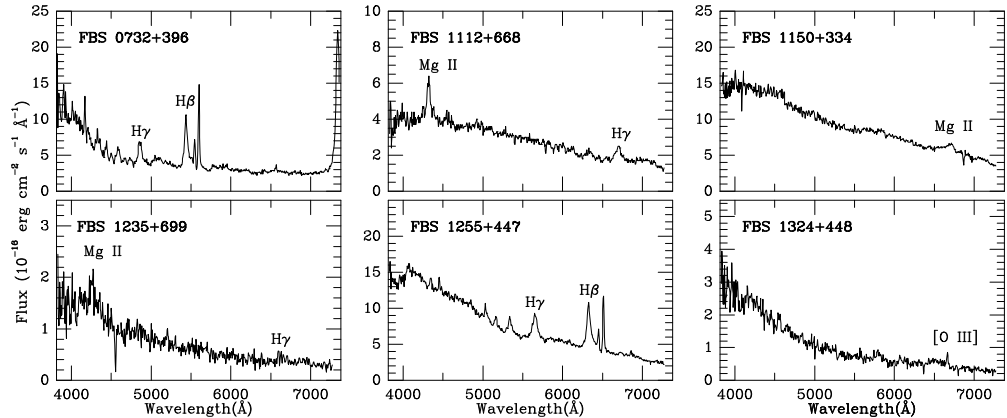}{5cm}{0}{105}{105}{-170}{-100}
\caption{Spectra of the six newly discovered QSOs.}
\end{figure}

The PG and FBS samples have about 2\,250 deg$^{2}$ in common. Out of the 
1\,103 FBS objects, 618 are within the PG fields, 276 being in the PG sample.
Thirty-six PG objects have not been found in the FBS. So 88\% (276/312) 
of the PG objects have been discovered. The 36 undiscovered objects have been 
examined on the FBS plates; 24 have a weak UV excess (the PG survey finds a 
significant fraction of stars with $U - B$ $\sim$ $-$0.4, while the FBS 
is relatively insensitive for $U - B$ $>$ $-$0.7). From this, we conclude 
that the FBS is $\sim$ 95\% complete for $U - B$ $<$ $-$0.5. This is 
significantly larger than the 67\% success rate obtained from the UBV stars; it 
is probably due to the fact that, in principle, the PG survey contains only 
objects brighter than $B$ = 16.2. 
There are 25 PG QSOs in the FBS area, 23 of them have been found, confirming 
that the FBS is very efficient in discovering bright QSOs. 

\vspace*{-0.7cm}

\begin{center}
\begin{table}
\begin{center} 
\caption{Newly discovered QSOs} 
\vspace{2mm} 
\begin{tabular}{p{2cm}cccllcrl}
\tableline 
 FBS name & Mag. &  $O$  &   & (1)    & (2)    &   &  $b$~~& ~~$z$   \\
\tableline
 0732+396 & 16.0 & 14.70 & X & 10     & 20     & N &  25.1 & 0.118   \\
 1112+668 & 17.0 & 16.53 & X & 10     & \,~4   & Y &  47.9 & 0.544   \\
 1150+334 & 16.2 & 16.30 & R & \,~0.8 & \,~1.2 & Y &  76.0 & 1.40    \\
 1235+699 & 17.9 & 17.96 & x & \,~5   & \,~4   & Y &  47.4 & 0.522 ? \\
 1255+447 & 16.5 & 16.48 & X & 10     & 13     & Y &  72.6 & 0.300   \\
 1324+448 & 17.0 & 18.09 & X & \,~8   & \,~9   & Y &  71.1 & 0.331 ? \\
\tableline
\tableline
\end{tabular}
\vspace*{3mm}
\begin{tabular}{p{11cm}}
\scriptsize  
X: in the RASS-BSC; x: in the WGACAT; R: in the NVSS; (1): error of 
the {\it ROSAT} or VLA position (in arcsec); (2): distance between the 
FBS and {\it ROSAT} or VLA positions (in arcsec); Y: in the PG area; 
N: not in the PG area; O: APS $O$ magnitudes.
\end{tabular}
\end{center} 
\end{table}
\end{center}
\vspace*{-0.8cm}

\subsection{The AGN content of the FBS}

Thirty-four FBS objects are listed as QSOs in the V\'eron-Cetty \& V\'eron 
(1998) catalogue. Two more have
been shown to be QSOs by Hagen et al. (1998) and six have been identified in 
the present paper. There are therefore 42 known QSOs in the FBS, 
41 being at high galactic latitude ($\vert b \vert$ $>$ 30$^{\rm o}$). 
At high galactic latitude, all FBS objects associated with a 
RASS-BSC source have been identified. Among them, there are 25 QSOs. 
As about 60\% of all PG QSOs are RASS sources, and assuming that this 
is true for the FBS QSOs, we should have a 
total of about 43 QSOs in the FBS catalogue. This suggests that the number 
of QSOs still to be found in the FBS catalogue is very small as 
42 are already known.

All 114 AGNs from the PG survey have been observed at 5 GHz with the VLA 
(Kellerman et al. 1989); 35 (30\%) have been detected with a flux 
density larger than 3 mJy. The same fraction (12/40) of the known FBS QSOs 
have been detected in the NVSS survey, confirming that the number of QSOs in 
the as yet spectroscopically unobserved FBS objects is small and probably 
cannot exceed 10, as the fraction of radio-detected QSOs would then 
drop below 25\% and be significantly lower than the corresponding fraction for
the PG survey.

\subsection{Completeness of the PG survey}

In principle the PG survey selected all objects with $U - B$ $<$ $-$0.46 
and brighter than $B \sim$ 16.2; however, the $U - B$ colour was 
measured with 
a relatively large error (0.24$^{\rm m}$ rms) which induced an incompleteness 
estimated at around 12\%. Moreover, in the interval 
0.6 $<$ $z$ $<$ 0.8, the strong Mg\,{\sc II} $\lambda$2800 emission 
line is in the $B$ filter which gives a much redder $U - B$ colour 
than for neighbouring redshifts, resulting into an incompleteness of 28\% in 
this range (Schmidt \& Green 1983). 

The catalogue of mean UBV data on stars (Mermilliod \& 
Mermilliod 1994) contains 
283 stars in the magnitude range 12.0 $< B <$ 16.5 and with 
$U - B$ $<$ $-$0.40 in the full 10\,714 deg$^{2}$ area of the PG 
survey; 190 are included in the PG 
catalogue. Twenty-four stars photoelectrically observed because they were in 
the PG catalogue have been ignored; the 
overall completeness of the PG survey is therefore 64\% (166/259). 67\% 
(162/241) of the stars brighter than $B$ = 16.2 were found in the PG survey, 
while only 22\% (4/18) of the weaker stars  were. The completeness of the 
PG survey rises from about 55\% at $U - B$ = $-$0.60 to 80\% at 
$U - B$ $<$ $-$1.0. For PG QSOs, the completeness should 
not be less than  $\sim$ 70\%. 

An independent estimate of the completeness of the PG survey comes from the 
comparison with the RASS-BSC. Of the 25 PG QSOs in the FBS 
area, 13 are brighter than $B$ = 16.16 and have been detected 
by {\it ROSAT}. Twelve of them have been 
found in the FBS. All FBS objects associated with a {\it ROSAT} source, located
in the PG area, have been spectroscopically identified; among them, there are 
eleven non-PG QSOs. No more than two are suspected 
to have been possibly bright enough to have been detected by the PG survey.
This suggests a completeness of about 87\% (13/15) with however a 
large uncertainty due to the smallness of the sample. 

Goldschmidt et al. (1992), K\"ohler et al.(1997) and La Franca \& 
Cristiani (1997) have claimed that the PG survey is incomplete by a factor of 
2 to 3, but their samples are quite small (5, 8 and 7 objects, respectively); in 
addition, we do not know if there is an offset between their magnitude scales 
and that of the PG survey (the mean differences between the PG and 
photoelectric $B$ magnitudes for 190 stars is equal to $-$0.02$^{\rm m}$). 
Therefore, their results should be considered as tentative.

\section{Building a ``complete'' QSO survey based on APS O magnitudes}

Because of their variability, it is an impossible task to compare directly two  
QSO surveys of the same region of the sky made at different epochs. However we
now have, for a large fraction of the sky, the possibility to extract from the
APS database, for any object, the $O$ magnitude as measured on the Palomar Sky 
Survey plates with an accuracy of about 0.2$^{\rm m}$ 
(Pennington et al. 1993). By doing this for all known QSOs found in 
the course of a 
number of different surveys of the same area of the sky, we may hope 
to reach as near as possible 
from an ideal survey complete to a well defined limiting magnitude. 

We have extracted the APS $O$ magnitudes, when available, for all objects in
the QSO catalogue (V\'eron-Cetty \& V\'eron 1998) brighter than $B$ = 17, 
with $M_{B}$ $<$ $-$24.0 and $z$ $<$ 2.15, located in the 2\,400 deg$^{2}$ 
of the FBS at $\vert b \vert$ $>$ 30$^{\rm o}$. Whenever this $O$ magnitude 
exists, we give it the preference. There are 18 such QSOs with $O <$ 16.0 
(from a sample of 105 UV-excess stars, we have found that 
$\rm \langle O - B \rangle$ = $-$0.17 with a rms error of 0.25$^{\rm m}$), and 
seven with $B <$ 16.2. 
Thus our ``complete'' sample contains between 18 and 25 QSOs brighter than
$B$ = 16.2, or 0.0075 to 0.010 deg$^{-2}$, i.e. 1.2 to 1.6 times larger than 
the PG surface density. It should be possible, when the APS database will 
be completed, to get the $O$ magnitudes of the seven objects for which 
they are not yet available. 

\vspace*{4mm}

\acknowledgments
A.\,M. Mickaelian is grateful to the CNRS for making possible his visit 
to OHP. A.\,C. Gon\c{c}alves acknowledges the {\it Funda\c{c}\~ao 
para a Ci\^encia e a Tecnologia}, Portugal (PRAXIS XXI/BD/5117/95 PhD. grant).

\vspace*{3mm}




\end{document}